# Reducing Spectral Oscillations for Robust Reference Frequency-Based Ultrasound Attenuation Estimation in Harmonic Imaging

U-Wai Lok†, Jingke Zhang†, Chengwu Huang, Tao Wu, Jieyang Jin, Ryan M. DeRuiter, Jingyi Yin, Lijie Huang, Yanzhe Zhao, Kaipeng Ji, Kate M. Knoll, Dawn Boynton, Kymberly D. Watt, Kathryn A. Robinson, Joshua D. Trzasko, Matthew Callstrom, and Shigao Chen*

*Abstract*— **Ultrasound attenuation coefficient estimation (ACE) has emerged as a quantitative imaging biomarker for noninvasive assessment of hepatic steatosis. A system-independent technique based on spectral normalization, known as the reference frequency method (RFM), was previously proposed to estimate ACE without requiring a well-calibrated reference phantom. Furthermore, incorporating harmonic imaging can significantly suppress reverberation signals. In previous clinical study, RFM has achieved high correlation with MRI-PDFF, demonstrating its potential for clinical application. However, a major challenge of RFM is the presence of oscillations in the frequency power-ratio decay curves (FPDCs), which can distort the linear fitting used to estimate the attenuation coefficient and consequently degrade ACE accuracy. These oscillations arise from constructive and destructive interference among backscattered echoes, resulting in oscillatory fluctuations in the measured power spectrum that propagate into the FPDCs. To address this limitation, we propose a transmission scheme combining multiple frequencies and steering angles to mitigate oscillations in the FPDCs. The hypothesis is that the varying transmission frequencies and steering angles produce distinct ultrasound signal patterns, resulting in weakly correlated oscillation patterns across the resulting FPDCs. Averaging these FPDCs suppresses the interference-induced oscillations while preserving the attenuation-dependent decay trend, thereby improving linearity and the accuracy of ACE results. In *in-vitro* experiments using calibrated phantoms (0.5 and 0.76 dB/cm/MHz) demonstrated that the proposed method improved FPDC linearity and ACE accuracy compared with conventional RFM, achieving an $R^2$ of 0.99 and attenuation coefficient estimates of 0.51 and 0.77 dB/cm/MHz, versus an $R^2$ of 0.89 and estimates of 0.56 and 0.70 dB/cm/MHz for conventional RFM. The proposed method also demonstrated superior performance in a pilot patient study (*n*=15), achieving a stronger correlation with magnetic resonance imaging proton density fat fraction (MRI-PDFF) (R = 0.89 vs. 0.83 for conventional RFM) while reducing inter-measurement variability, indicating improved robustness and clinical potential.**

*Index Terms*—**Ultrasound attenuation coefficient estimation, harmonic imaging, Oscillation reduction, Fatty liver detection**

## I. Introduction

INCREASED hepatic fat content is a hallmark of hepatic steatosis, which may progress to fibrosis, cirrhosis, liver failure, or hepatocellular carcinoma. Compared with liver biopsy, the clinical gold standard, proton density fat fraction (PDFF) [1] obtained via magnetic resonance imaging (MRI) is a noninvasive alternative and is frequently used as a reference standard for liver fat quantification. However, MRI can be limited by low accessibility, which restricts its use for frequent monitoring. Ultrasound attenuation coefficient estimation (ACE) has demonstrated the potential for quantifying fat content in the human liver. Previous studies have shown that pathological processes alter the median ACE in proportion to the accumulation of fatty droplets [2, 3]. Consequently, there is a strong demand for low-cost, widely accessible, and accurate ACE methods for reliable quantification of liver fat.

The feasibility of ultrasound ACE has been demonstrated using several methodological approaches. Spectral shift techniques [4, 5] which estimate attenuation by analyzing the downshift in the mean frequency or by comparing power spectra before and after tissue propagation and notably do not require a reference phantom. In contrast, reference phantom-based methods, such as spectral difference [6], spectral log-difference [7], and hybrid approaches [8], use a calibrated phantom to compensate for system-dependent effects (e.g., focusing, diffraction, and time-gain compensation), improving robustness but introducing practical challenges related to phantom maintenance and stability. Another group of methods employs least-mean-square error filtering [9] to enhance estimation performance. Additionally, model-based and multiparameter approaches [10, 11] have been proposed that estimate multiple parameters (such as attenuation, and/or backscatter coefficient) simultaneously. However, the substantial computational cost of these approaches may limit their real-time implementation. More recently, machine learning-based approaches have been presented, with the use of attenuation as a parameter or the use of deep-learning neural networks with data-driven regression models [12, 13]. While

†U-Wai Lok and Jingke Zhang equally contributed to this work

U-Wai Lok, Jingke Zhang, Chengwu Huang, Tao Wu, Jieyang Jin, Ryan M. DeRuiter, Jingyi Yin, Lijie Huang, Yanzhe Zhao, Kaipeng Ji, Kate M. Knoll, Dawn Boynton, Kathryn A. Robinson, Joshua D. Trzasko, Matthew Callstrom, and Shigao Chen are with the Department of Radiology, Mayo Clinic, College of Medicine and Science, Rochester, MN, USA (correspondence e-mail: Chen.Shigao@mayo.edu).

Kymberly D. Watt is with the Division of Gastroenterology, Mayo Clinic College of Medicine and Science, Rochester, MN, 55905 USA.

The study was supported partially by the National Institutes of Health under award numbers of R01DK127978. The content is solely the responsibility of the authors and does not necessarily represent the official views of the National Institutes of Health. The Mayo Clinic and some of the authors (U.L,, J.Z., and S.C.) have a potential financial interest (Patents/Licensing) related to the technology referenced in the research.

these methods offer fast and calibration-free estimation, they require large, annotated datasets and often sacrifice interpretability compared to traditional techniques.

Recently, a system-independent and computationally efficient reference frequency method (RFM) was proposed for estimating ultrasound attenuation coefficients without requiring a well-calibrated reference phantom. System-dependent parameters are canceled through spectral normalization using adjacent ultrasound frequency components, enabling the method to be applied across various transducers (e.g., linear or curved arrays) and beam patterns (e.g., focused or unfocused). Furthermore, combining RFM with harmonic imaging further improved performance in the presence of reverberation clutter [14, 15]. Furthermore, compared with the reference phantom method, the reference frequency method is generally more robust to slight sound speed mismatch because attenuation is estimated from the relative spectral decay within the same tissue rather than by comparison with a reference phantom.

In ideal conditions, this frequency power ratio decay curve (FPDC) yields a smooth, monotonic function; however, in practice, the FPDC often exhibits oscillations. As the attenuation coefficient is derived by linear fitting of the FPDC, the oscillations in FPDC will compromise the accuracy of ACE. These oscillations may arise from: (i) background noise, (ii) structural heterogeneity in *in vivo* tissues (non-uniform structure), and (iii) spatial fluctuations due to constructive and destructive interference in backscattered signals. Our previous work [16] proposed methods to suppress noise contamination in the signals used for ACE, thereby improving estimation robustness. Additionally, non-uniform structure detection and removal method [17] has been proposed to further improve the robustness of ACE by addressing the oscillations caused by non-uniform structures in the tissue. However, the random distribution of tissue scatterers gives rise to constructive and destructive interference among backscattered echoes, producing oscillatory fluctuations in the local power spectrum. These fluctuations lead to large oscillations in the FPDCs and consequently unstable ACE estimates. We hypothesize that FPDCs acquired using different transmission frequencies and steering angles are weakly correlated because the corresponding transmissions would produce distinct ultrasound signal patterns. By averaging multiple low-correlated FPDCs, the overall oscillation of the averaged FPDC can be reduced, thereby improving linearity and the final ACE results. The accuracy of the proposed method was first evaluated with two calibrated tissue-mimicking phantoms. The *in vivo* feasibility of the proposed method was tested on 15 patients who underwent clinically indicated MRI of the liver. PDFF acquired with MRI was used as the reference standard [18]. The acquired attenuation coefficient was correlated with MRI-PDFF to evaluate the performance of the ACE with the conventional RFM and the proposed RFM approach.

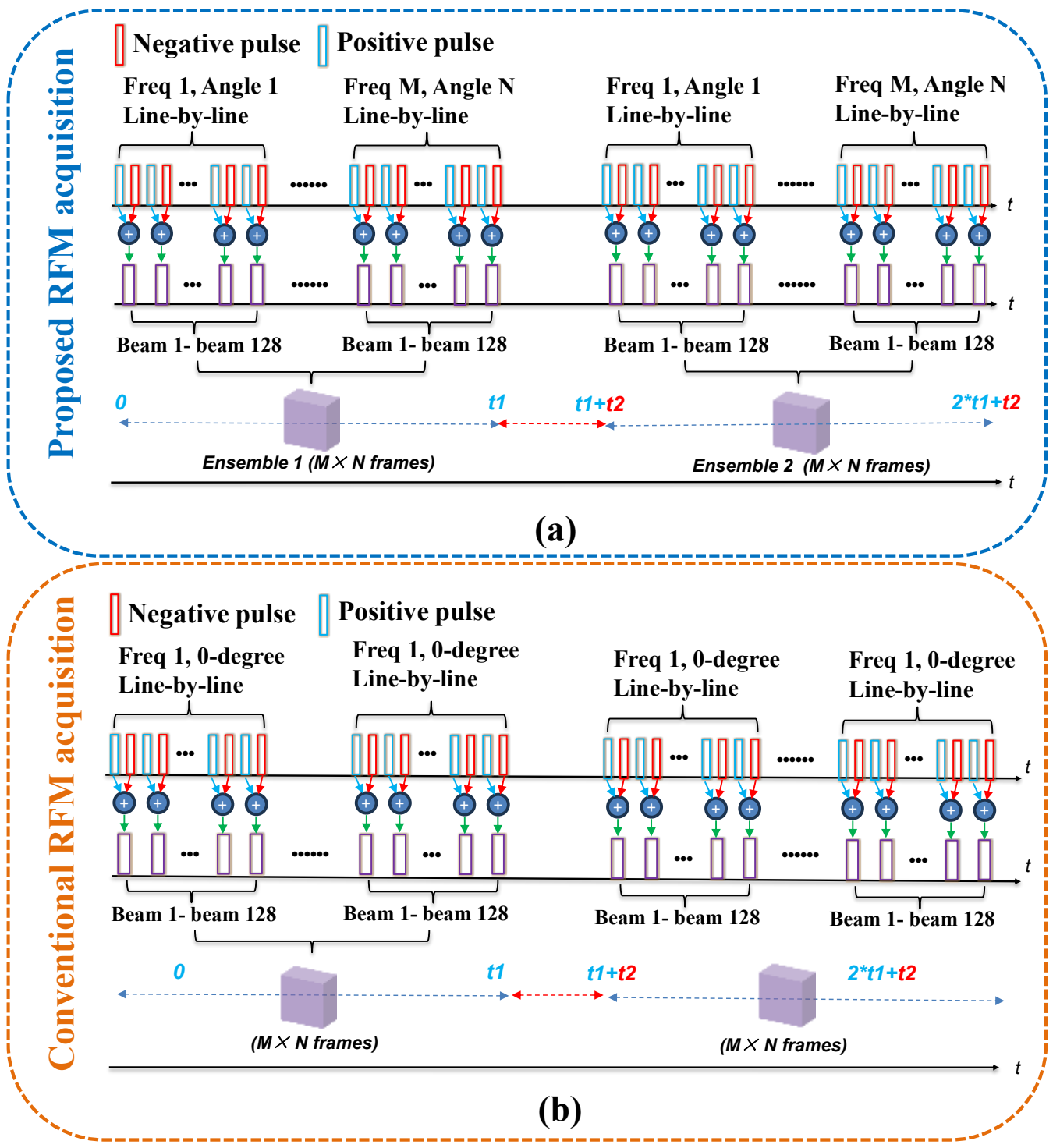

**Fig. 1.** Schematic illustration of data acquisition sequence using pulse inversion for (a) proposed RFM, and (b) conventional RFM methods. In the proposed sequence, each ensemble consists of acquisitions using $M$ transmission frequencies and $N$ steering angles, whereas the conventional sequence forms each ensemble by repeating acquisitions with the same transmission frequency and steering angle $M \times N$ times.

## II. Methods and Materials

### A. Principles of RFM-based ACE

In this study, the power spectrum of the backscattered ultrasound signals $S(f_i, z_k)$ can be expressed as [19]

$$S(f_i, z_k) = G(f_i) \cdot T(z_k) \cdot D(f_i, z_k) \cdot BC(f_i) \cdot A(f_i, z_k) \quad (1)$$

, where $G(f_i)$ represents the transmit and receive transducer responses at frequency $f_i$ ($i$ is the frequency component index); $T(z_k)$ is the time gain compensation and $D(f_i, z_k)$ is the combined effects of focusing, beamforming, and diffraction at depth $z_k$; $BC(f_i)$ indicates the backscatter coefficient and $A(f_i, z_k)$ is the frequency-dependent attenuation defined as [19]

$$A(f_i, z_k) = \exp(-4a f_i \, z_k) \quad (2)$$

, where $a$ is the frequency-dependent ultrasound attenuation coefficient. By assuming the differences in beamforming and diffraction effects between two adjacent frequency components (i.e., $f_i$ and $f_{i-1}$) are negligible, the power spectrum can be normalized by calculating the power ratio $R(f_i, z_k)$ between adjacent frequency components $S(f_i, z_k)$ and $S(f_{i-1}, z_k)$ to cancel the effects of TGC and diffraction, which can be expressed as [20]

$$R(f_i, z_k) = \frac{S(f_i, z_k)}{S(f_{i-1}, z_k)} = \frac{G(f_i) \cdot BC(f_i) \cdot A(f_i, z_k)}{G(f_{i-1}) \cdot BC(f_{i-1}) \cdot A(f_{i-1}, z_k)} \quad (3)$$

. After taking the natural logarithm on both sides of (3), we obtain the following linear relationship between frequency power ratio ($\ln[R(fi, z_k)]$) and imaging depth ($z_k$):

$$\ln[R(f_i, z_k)] = \ln[G(f_i)] - \ln[G(f_{i-1})] + \ln[BC(f_i)] - \ln[BC(f_{i-1})] - 4a(f_i - f_{i-1})z_k \quad (4)$$

. Then the attenuation coefficient can be estimated from the slope of the FPDC with respect to each frequency component [20].

### *B. Conventional and proposed RFMs*

Fig. 1(a) shows the data acquisition of the proposed RFM; pulse-inversion (PI) harmonic imaging [21, 22] was performed using a conventional line-by-line focused acquisition with 128 transmit beams. Specifically, each PI transmit event consisted of two consecutive transmissions with opposite polarities, and the corresponding RF channel data were accumulated and stored in memory. In addition, multiple transmit harmonic frequencies (Freq 1 to Freq M in Fig. 1a) and steering angles (Angle 1 to Angle N in Fig. 1a) were employed, resulting in a total of M×N transmit frequency-angle combinations. In this study, a data ensemble was defined as the complete set of RF channel data acquired using all M transmit frequencies and N steering angles. An example of two consecutive data ensembles is illustrated at the bottom of Fig. 1(a), with a total acquisition time of $2 \times t_1 + t_2$, where $t_1$ represents the acquisition time for the first data ensemble and $t_2$ represents the time interval between the acquisitions of the first and second data ensembles. For fair comparison, the total number of transmissions and acquisition time of the conventional RFM were matched to those of the proposed RFM, as shown in Fig. 1(b). For the conventional RFM acquisition, non-steered transmissions at a single harmonic frequency of 4 MHz were repeated to acquire a total of M×N frames for each data ensemble. This ensured that any performance improvement could be attributed to the proposed multi-frequency and multi-angle transmission strategy, rather than to the increased number of acquisition events.

As illustrated in Fig. 2, following data acquisition (using three transmit frequencies and steering angles as an example in step 1), a customized baseband beamforming (step 2 in Fig. 2) was applied to reconstruct the in-phase/quadrature (IQ) data with a pixel size of 0.5 λ (lateral) by 0.25 λ (axial), where λ denotes the wavelength. In addition, the beamforming grid for each steered transmission was defined along the beam direction

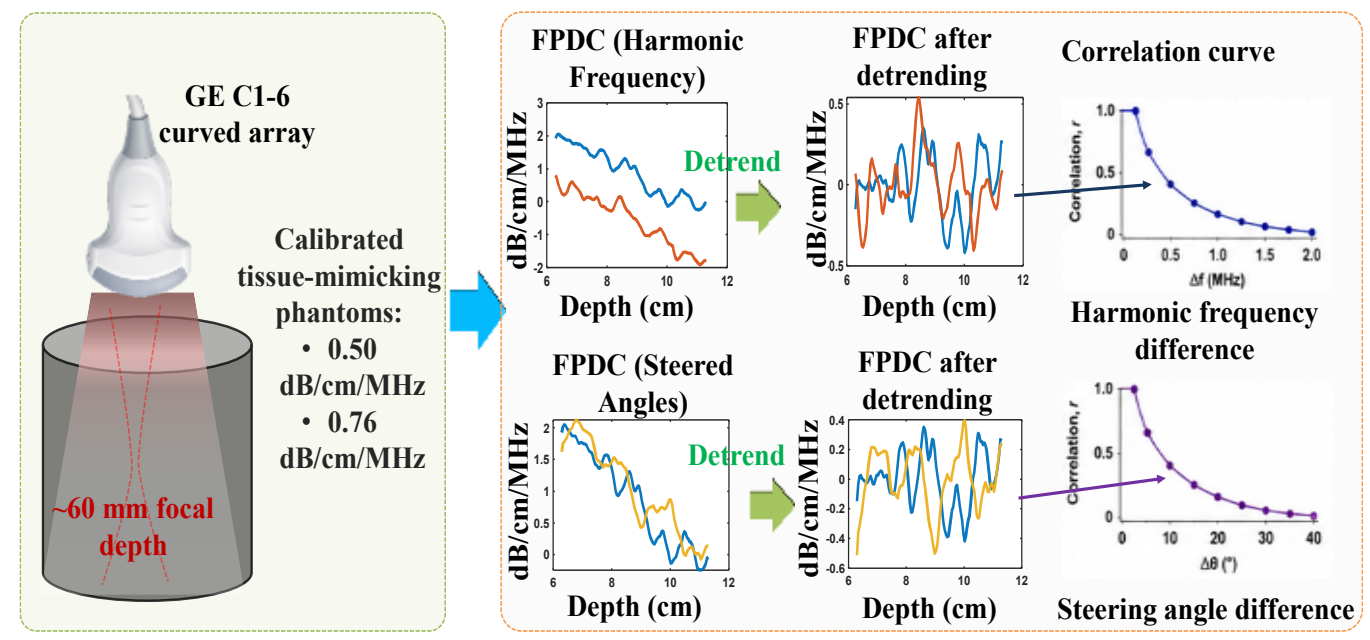

**Fig. 3.** Illustration diagram of obtaining correlations between the frequency power ratio decay curves as a function of transmitting frequency increments and steering angle increments.

to ensure that FPDCs were estimated with respect to the true propagation distance of each transmitted beam. A region-of-interest (ROI) was selected (step 3 in Fig. 2) and divided into multiple blocks corresponding to different sub-ROIs (step 4 in Fig. 2). The blocks are shown as separated regions for visualization only; the overlap between adjacent blocks can be adjusted according to the ROI size. In our implementation, 50-60 blocks with a 10-pixel step size in both the lateral and axial directions were used. Power spectra (step 5 in Fig. 2) and the corresponding FPDCs (step 6 in Fig. 2) were computed independently for each acquisition at different transmit frequencies and steering angles for each block following the conventional RFM processing [14]. Subsequently, the resulting FPDCs within each block (step 6 in Fig. 2) were averaged across all transmissions, yielding a smoother overall FPDC (blue curve in step 7 in Fig. 2). The linear fitting is then performed on the averaged FPDC (red-dotted line in step 7 of Fig. 2), with the fitted slope taken as the ACE value and assigned to the corresponding block (block 1 in step 8 of Fig. 2). Finally, ACE map was overlayed onto the corresponding B-mode image for displaying. All post-processing was conducted on a workstation with Intel Xeon(R) W-2265 3.5 GHz CPU and

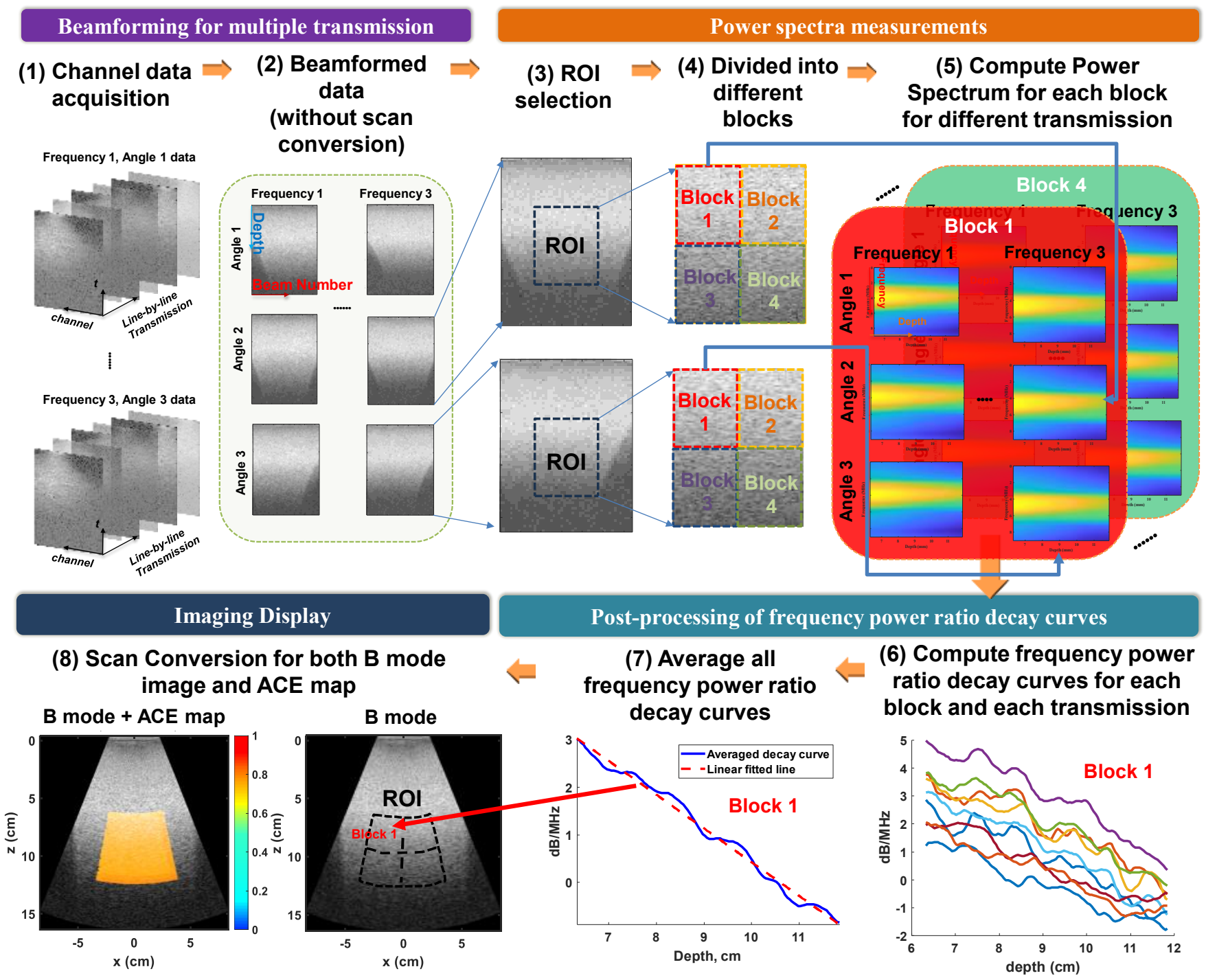

**Fig. 2.** Schematic illustration of the post-processing pipeline, including beamforming process, power spectra analysis and FPDC computation for multiple angles and frequencies to improve estimation stability, averaging FPDCs, and generating final ACE map for display.

512 GB RAM using MATLAB 2021a (MathWorks, Natick, MA, USA). The computational time for beamforming was approximately 45.7 s per transmission, while the calculation of each FPDC required approximately 2.3 s.

### *C. Phantom study*

In this study, RF channel data were acquired using a C1-6D curved-array transducer (1-6 MHz, GE Healthcare, Wauwatosa, WI, USA) connected to a Vantage 256 ultrasound research system (Verasonics Inc., Kirkland, WA, USA). The transmitted focal depth was set to 60 mm and the sampling rate was approximately 18 MHz.

The proposed method was validated in two calibrated tissue-mimicking phantoms with attenuation of 0.50 and 0.76 dB/cm/MHz (Sun Nuclear, Melbourne, FL) and compared to conventional RFM method [14]. Our first objective was to determine the optimal transmission frequency and steering angle settings by evaluating the FPDC correlation as a function of frequency separation ($\Delta f$) and steering angle separation ($\Delta\theta$), as lower correlation between FPDCs is expected to improve oscillation suppression through averaging. As illustrated in Fig. 3, the reference frequency and steering angle were defined as $\Delta f = 0$ and $\Delta\theta = 0$, respectively. The step sizes used to calculate the correlation coefficients for transmission frequency and steering angle were 0.1 MHz and 1°, respectively. The normalized correlation coefficient for each $\Delta f$ or $\Delta\theta$ was calculated between the detrended FPDC obtained at each increment and the detrended FPDC obtained at the reference frequency and steering angle. Detrending was performed using MATLAB's *detrend* function, and normalized cross-correlation was subsequently calculated using MATLAB's *xcorr* function. The resulting correlation coefficients for different frequency and steering angle increments were used to generate correlation curves as a function of the corresponding increments. These correlation curves were then used to determine the optimal thresholds for selecting transmission frequency and steering angle parameters.

The second objective of the phantom study was to evaluate the robustness of the proposed method in the presence and absence of phase aberration, using the transmission frequencies and steering angles selected in the first objective. Two tissue-mimicking phantoms with calibrated attenuation coefficients of 0.50 and 0.76 dB/cm/MHz were used as reference targets for ACE. Additionally, the numbers of transmit frequencies and steering angles were both set to 3 ($M = 3$ and $N = 3$ in Fig. 1), resulting in an acquisition time of 61.4 ms for each data ensemble, with a 0.6 ms interval between consecutive ensembles. For the phantom study, one data ensemble was acquired for each measurement. To simulate the phase aberration encountered in clinical imaging, a layer of pork belly approximately 17 mm thick was placed directly on the surface of each phantom, with ultrasound gel applied at the interfaces to ensure adequate acoustic coupling. The ACE obtained with and without the pork belly were evaluated and compared.

### *D. In vivo human liver study*

For clinical validation, the proposed method was evaluated in 15 patients. This study was approved by the Institutional Review Board of the Mayo Clinic, and written informed consent was obtained from each participant at the time of enrollment. All participants fasted for more than 6 hours prior to imaging. MRI-PDFF was measured using a GE Optima 450 MRI scanner (GE Healthcare) with the IDEAL IQ sequence [35]. The same Verasonics acquisition sequence as that used in the phantom study was employed for *in vivo* data acquisition. The pulse repetition frequency was set to 416.67 Hz. Additionally, the numbers of transmit frequencies and steering angles were both set to 3 ($M = 3$ and $N = 3$ in Fig. 1). These settings resulted in an acquisition time of 61.4 ms for each ensemble, with a 0.6 ms interval between consecutive ensembles, corresponding to an imaging frame rate of approximately 16 Hz. Two data ensembles were acquired for each patient. Ten consecutive measurements were collected from each patient.

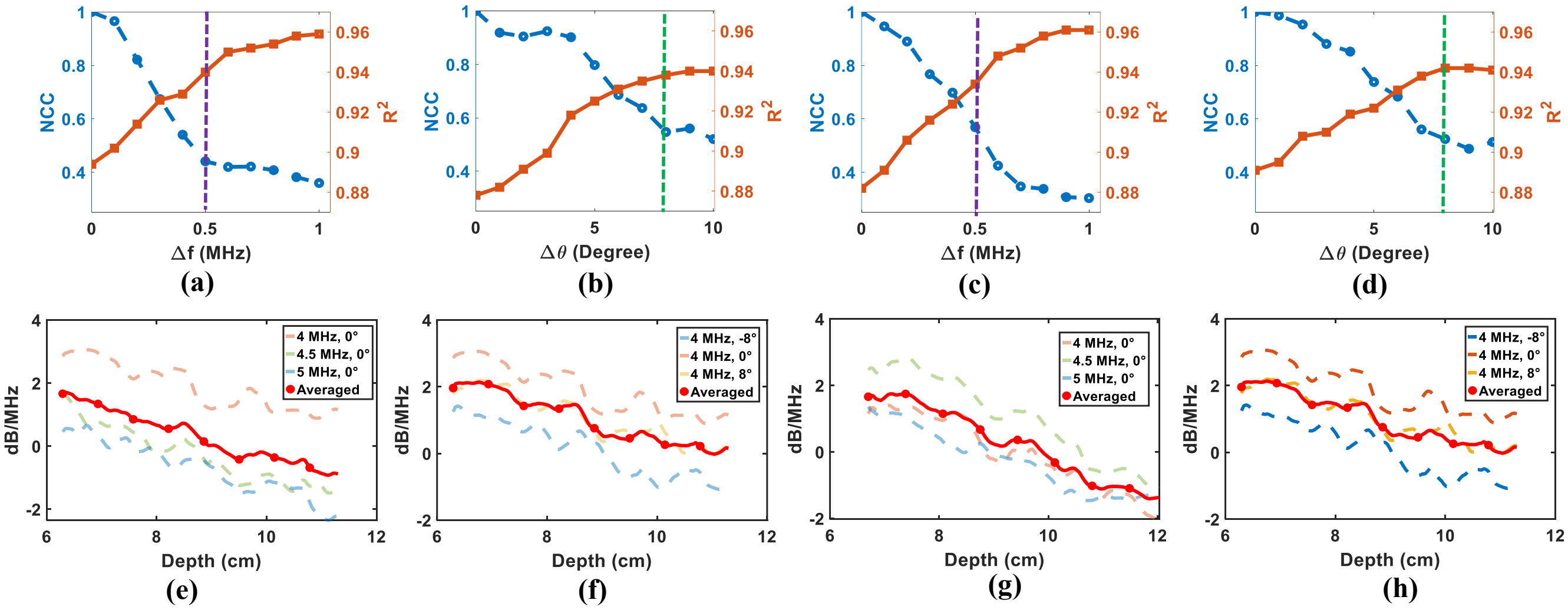

**Fig. 4.** (a-d) Normalized cross-correlation and coefficient of determination ($R^2$) between FPDCs as functions of harmonic frequency increment ($\Delta f$) and steering angle increment ($\Delta\theta$). Results for the 0.5 dB/cm/MHz phantom are shown in (a) and (b), while results for the 0.76 dB/cm/MHz phantom are shown in (c) and (d). FPDCs correspond to three distinct harmonic frequencies (light-colored dotted lines) and their average (deep-colored red solid line) are shown in (e) and (g) for the 0.5 and 0.76 dB/cm/MHz phantoms, respectively. FPDCs correspond to three distinct steering angles (light-colored dotted lines) and their average (deep-colored red solid line) are shown in (f) and (h) for the 0.5 and 0.76 dB/cm/MHz phantoms, respectively.

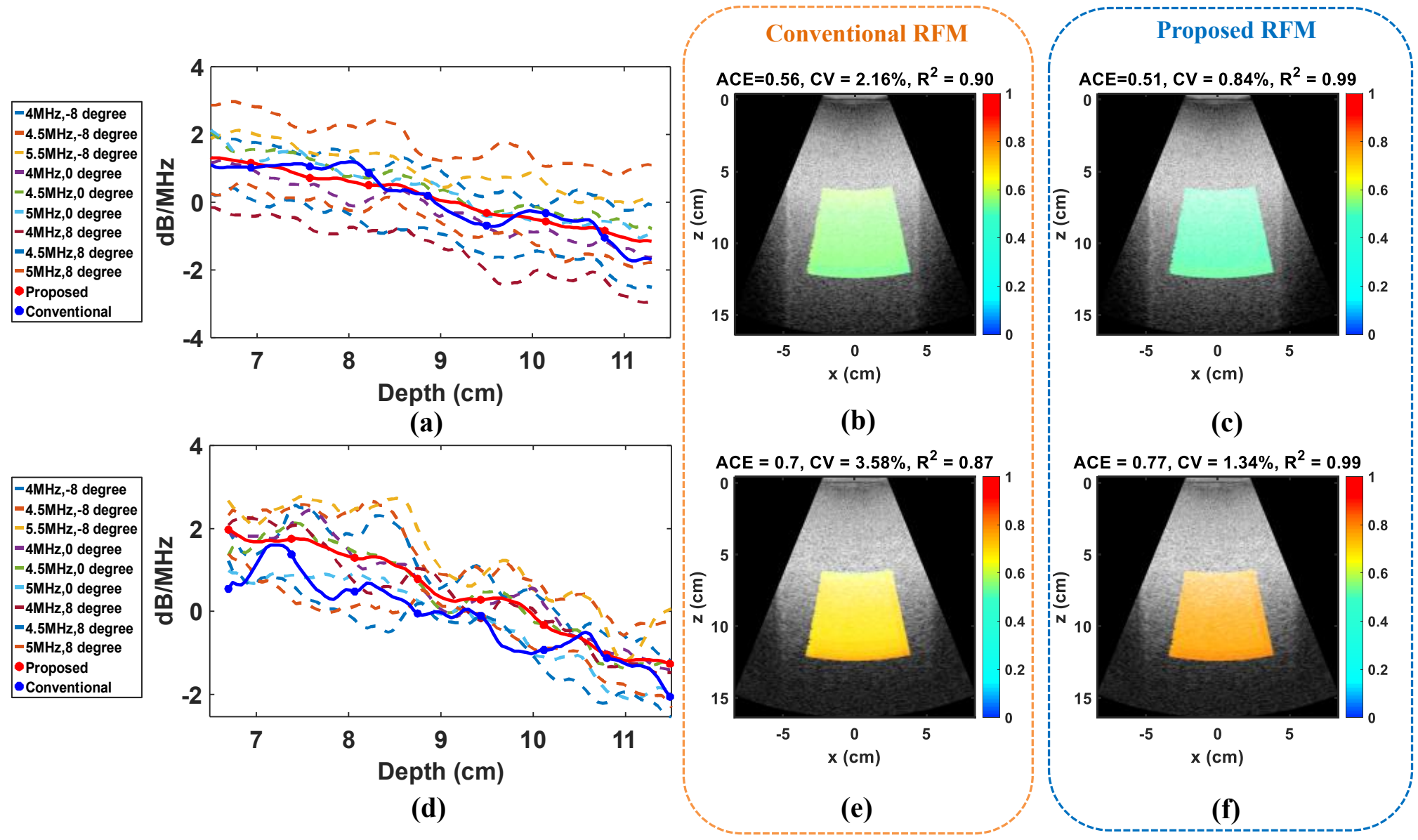

**Fig. 5.** (a) Frequency power ratio decay curve with different harmonics frequencies and steering angle for 0.5 dB/cm/MHz phantom, and the corresponding ACE results using (b) the conventional and proposed RFM methods and (c) the proposed RFM approach. (d) Frequency power ratio decay curve with different harmonics frequencies and steering angle for 0.76 dB/cm/MHz phantom, and the corresponding ACE results using (e) the conventional and proposed RFM methods and (f) the proposed RFM approach. The unit of the color bar in (b), (c), (e) and (f) is dB/cm/MHz.

### *E. Evaluation metrics*

To evaluate the correlation of FPDCs for different frequency and angular separations, the normalized cross-correlation (NCC) was calculated between the detrended FPDC at each separation and the corresponding detrended FPDC obtained under the reference condition ($\Delta f = 0$ or $\Delta\theta = 0$). A lower NCC value is desirable because it indicates reduced similarity between the oscillatory components of different FPDCs, enabling destructive interference during averaging and resulting in a smoother FPDC curve with improved linearity for attenuation coefficient estimation.

To investigate the performance of ACE estimation in the phantom study, the mean ACE and the coefficient of variation (CV) were evaluated. The mean ACE was obtained by averaging the estimated ACE values obtained from all blocks (step 4 of Fig. 2) and compared with the calibrated phantom value to assess the estimation accuracy of the conventional RFM and proposed RFM methods. The CV was calculated as the standard deviation of ACE values across all blocks divided by the mean ACE value and expressed as a percentage:

$$CV = \frac{Standard\ deviation\ (ACE)}{mean\ ACE} \times 100\% \tag{5}$$

. A lower CV indicates reduced spatial variability among blocks and improved consistency of the ACE estimates.

Additionally, the median coefficient of determination (median $R^2$) was used to quantify the linearity and oscillation characteristics of the FPDCs. The individual $R^2$ value was obtained by fitting a linear regression model to a FPDC within a block, and the median $R^2$ was calculated by taking the median value of $R^2$ values from FPDCs across all inter-blocks. A higher median $R^2$ value is desirable because it indicates improved linearity of the FPDCs and reduced influence of oscillatory components.

Furthermore, for the *in vivo* study, the median of the mean ACE values obtained from the ten consecutive measurements was used for analysis, consistent with previous work [34]. The median ACE values were correlated with the corresponding MRI proton density fat fraction (MRI-PDFF) measurements using the MATLAB *corrcoef* function to evaluate the performance of both the proposed RFM and the conventional RFM. In addition, the pooled standard deviation (pooled SD) was assessed. The pooled standard deviation was calculated as the square root of the weighted average of the within-patient variances, where the degrees of freedom of each patient were used as weighting factors:

$$Pooled\ SD = \frac{\sqrt{\sum_{i=1}^{K}(n_i-1)SD_i^2}}{\sqrt{\sum_{i=1}^{K}(n_i-1)}} \tag{6}$$

, where $SD_i$, $n_i$ and $K$ represent the standard deviation of the repeated measurements for the $i$ th patient, the number of measurements per patient, and the total number of patients, respectively. A lower pooled SD indicates reduced inter-acquisition variability and improved repeatability, reflecting more consistent ACE estimates from different patients' measurements.

## III. Results

### *A. Correlation between FPDCs obtained using different transmitting frequencies and steering angles*

Normalized cross-correlations (NCCs) between the detrended FPDCs and the measurement harmonic frequency interval ($\Delta f$) were evaluated for both 0.5 and 0.76 dB/cm/MHz tissue-mimicking phantoms, as shown in Figs. 4(a) and 4(c). The NCC and coefficient of determination ($R^2$) are represented by the blue and red curves, respectively. For both phantoms, the NCC decreased as $\Delta f$ increased, whereas the corresponding $R^2$ increased with increasing $\Delta f$.

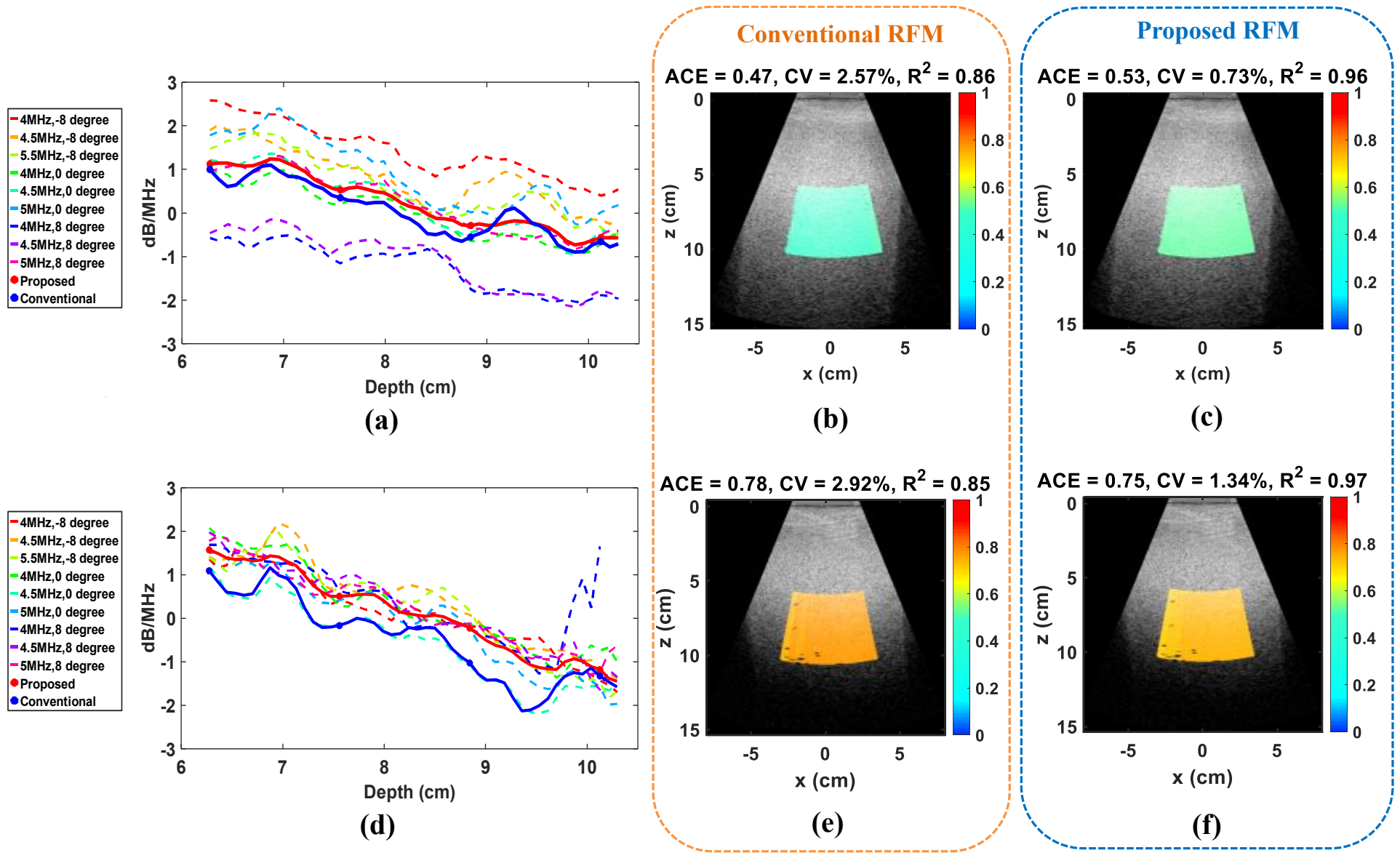

**Fig. 6.** (a) Frequency power ratio decay curve with different harmonics frequencies and steering angle for 0.5 dB/cm/MHz phantom with interposed pork belly, and the corresponding ACE results using (b) the conventional and proposed RFM methods and (c) the proposed RFM approach. (d) Frequency power ratio decay curve with different harmonics frequencies and steering angle for 0.76 dB/cm/MHz phantom with pork belly placed above it, and the corresponding ACE results using (e) the conventional and proposed RFM methods and (f) the proposed RFM approach. The unit of the color bar in (b), (c), (e) and (f) is dB/cm/MHz.

Similarly, the NCC decreased as the steering angle interval ($\Delta\theta$) increased for both 0.5 and 0.76 dB/cm/MHz phantoms, as shown in Figs. 4(b) and 4(d), while the corresponding $R^2$ values increased with increasing $\Delta\theta$ for both phantoms.

To balance the tradeoff between $R^2$, NCC, and the available transducer bandwidth (1.8-5 MHz), three harmonic frequencies with a frequency interval of $\Delta f$ = 0.5 MHz (indicated by the purple dotted lines in Figs. 4(a) and 4(c)) and three steering angles with an angular interval of $\Delta\theta$ = 8° (indicated by the green dotted lines in Figs. 4(b) and 4(d)) were selected for subsequent ACE evaluation.

Figure 4(e) shows the FPDCs obtained from three harmonic frequencies (4.0, 4.5, and 5.0 MHz; dotted lines) and their average (red solid line) for the 0.5 dB/cm/MHz phantom. The averaged FPDC exhibited better linearity ($R^2$ = 0.97) compared with the individual FPDCs ($R^2$ = 0.86, 0.88, and 0.87 for 4.0, 4.5, and 5.0 MHz, respectively). Figure 4(f) shows the FPDCs obtained from three steering angles (−8°, 0°, and 8°) at a harmonic frequency of 4.0 MHz, together with their averaged FPDC (red solid line) for the 0.5 dB/cm/MHz phantom. The averaged FPDC also demonstrated improved linearity ($R^2$ = 0.95) compared with the individual FPDCs ($R^2$ = 0.87, 0.86, and 0.87 for −8°, 0°, and 8°, respectively).

Similarly, Fig. 4(g) presents the FPDCs obtained from three harmonic frequencies (4.0, 4.5, and 5.0 MHz; dotted lines) and their average (red solid line) for the 0.76 dB/cm/MHz phantom. The averaged FPDC achieved a higher linearity ($R^2$ = 0.98) than the individual FPDCs ($R^2$ = 0.89, 0.90, and 0.88 for 4.0, 4.5, and 5.0 MHz, respectively). Figure 4(h) shows the FPDCs obtained from three steering angles (−8°, 0°, and 8°) at a harmonic frequency of 4.0 MHz, along with their averaged FPDC (red solid line) for the 0.76 dB/cm/MHz phantom. The averaged FPDC again exhibited better linearity ($R^2$ = 0.95) than the individual FPDCs ($R^2$ = 0.88, 0.89, and 0.91 for −8°, 0°, and 8°, respectively). In both phantoms, averaging the three FPDCs obtained from different harmonic frequencies or steering angles produced a smoother FPDC with improved linearity.

### B. *Performance evaluation using attenuation phantoms*

Figure 5(a) shows the FPDCs obtained from the nine harmonic frequency and steering angle combinations (dotted lines), their averaged FPDC (red solid line), and the FPDC obtained using the conventional RFM method (blue solid line) for the 0.5 dB/cm/MHz phantom. The averaged FPDC was obtained from one data ensemble (all nine transmissions, with harmonic frequencies of 4.0, 4.5, and 5.0 MHz and steering angles of −8°, 0°, and 8°). Averaging across all decay curves yields a smoother and more stable curve with reduced oscillations compared with conventional RFM. The median $R^2$ values of all blocks demonstrate improved linearity (median $R^2$ = 0.99) with the proposed RFM (Fig. 5c) as compared to the conventional RFM (median $R^2$ = 0.90; Fig. 5b). Furthermore, the mean ACE value (0.51 dB/cm/MHz) between inter-blocks with the proposed method is closer to the calibrated value (0.5 dB/cm/MHz) of the phantom as compared to the conventional RFM method (mean ACE = 0.56 dB/cm/MHz). The CV of the proposed method between blocks (0.84%) using the proposed method is smaller than that of conventional RFM method (2.16%). Similar improvements are observed for the 0.76 dB/cm/MHz phantom. The FPDC obtained with the proposed RFM is smoother than that of the conventional RFM (Fig. 5(d)), and the linearity of the proposed RFM (Fig. 5(f), median $R^2$ = 0.99) outperforms that of the conventional RFM (Fig. 5(e), median $R^2$ = 0.87). Furthermore, the mean ACE value (0.77 dB/cm/MHz) between blocks with the proposed method is closer to the calibrated value of the phantom (0.76 dB/cm/MHz) as compared to the conventional RFM method (0.71 dB/cm/MHz). The CV of the proposed method between inter-blocks (1.34%) using the proposed method is smaller than that of conventional RFM method (3.58%). The proposed method is less sensitive to fitting-range selection and therefore more robust.

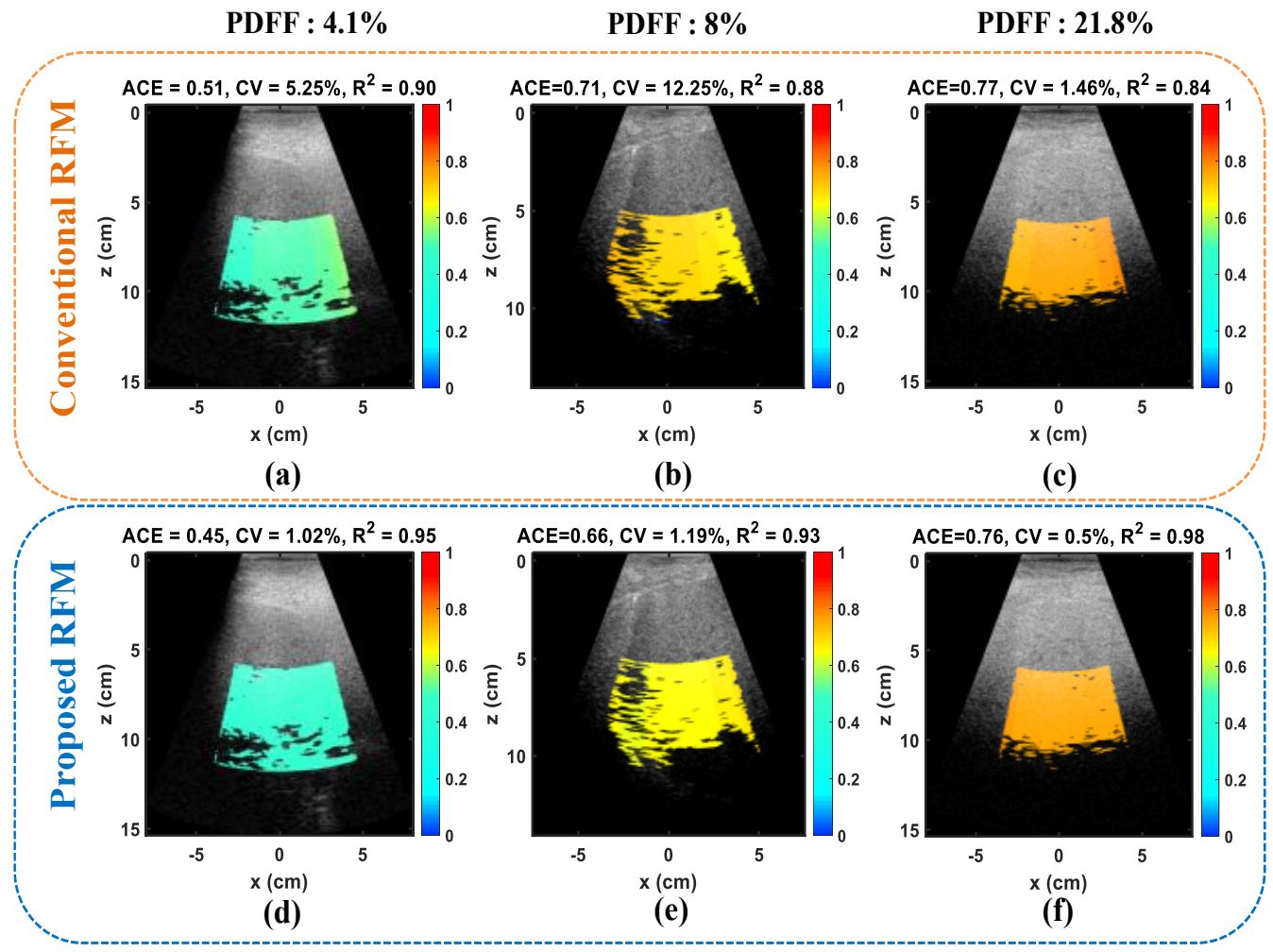

**Fig. 7.** Attenuation coefficient maps acquired from four patients with different proton density fat fraction (PDFF). (a,d): Attenuation image acquired from a patient with a PDFF of 4.1% using conventional and the proposed RFM method, (b,e): Attenuation image acquired from a patient with a PDFF of 8% using conventional and the proposed RFM method. (c,f): Attenuation image acquired from a patient with a PDFF of 21.8% using conventional and proposed RFM method.

Figure. 6(a) shows the FPDCs obtained from different harmonic frequencies and steering angles (dotted lines), their average of 9 FPDCs (red solid line) using the proposed RFM, and the FPDC (blue solid line) obtained using the conventional RFM method for a 0.5 dB/cm/MHz phantom with pork belly placed above it. The median $R^2$ values (Figs. 6(b) and 6(c)) further demonstrate improved linearity with the proposed RFM, whereas the conventional RFM exhibits lower linearity (median $R^2 = 0.96$ vs. 0.86). The CV of the proposed method between blocks (0.73%) using the proposed method is smaller than that of conventional RFM method (2.57%). Similar improvements are observed for the 0.76 dB/cm/MHz phantom. The FPDC obtained with the proposed RFM is smoother than that of the conventional RFM (Fig. 6(d)), and the linearity (Fig. 6(f), median $R^2 = 0.97$) with the proposed method outperforms that of the conventional RFM (Fig. 6(e), median $R^2 = 0.85$). The CV of the proposed method between blocks (1.34%) using the proposed method is smaller than that of conventional RFM method (2.92%).

### *C. Performance evaluation using in vivo liver data*

Figures 7(a)-(c) show the attenuation coefficient maps reconstructed using the conventional RFM method for patients with MRI-PDFF values of 4.1%, 8%, and 21.8%, respectively. Figures 7(d)-(f) show the corresponding ACE maps reconstructed using the proposed RFM method for the same patients. For both methods, the maps were reconstructed from two data ensembles comprising 18 frames in total. The mean ACE values between blocks using the conventional RFM and with the proposed method are near, around 0.01-0.06 dB/cm/MHz difference only. However, the CV with the proposed method in different blocks improved with the proposed method, the CVs are around 1.02%, 1.19% and 0.5% with MRI-PDFF values of 4.1%, 8%, and 21.8%, respectively. In contrast, the conventional method showed higher CVs, as 5.25%, 12.25% and 1.46%, respectively. Additionally, median $R^2$ values demonstrate improved linearity with the proposed method, yielding median $R^2$ values of 0.95, 0.93, and 0.98 for patients with MRI-PDFF values of 4.1%, 8%, and 21.8%, respectively. In contrast, the conventional method exhibited lower linearity, with median $R^2$ values of 0.90, 0.88, and 0.84.

Figures 8 (a) and (b) demonstrate the correlation plot of MRI-PDFF versus the median ACE values, where the correlation coefficient (R) evaluates the relationship between the two metrics. The proposed method improved the correlation with PDFF (R = 0.89 in Fig. 8b) compared to the conventional RFM method (R = 0.83 in Fig. 8a), suggesting that it provides ACE estimates that better reflect the degree of hepatic steatosis. In addition to the improved correlation, the proposed approach demonstrated enhanced stability of ACE estimation across repeated measurements. The proposed RFM method reduced the variation in ACE values, as indicated by the interquartile range (IQR). Furthermore, the pooled SD of ACE value was reduced from 0.0667 using the conventional method to 0.0378 using the proposed method across the 15 patients, indicating improved robustness and repeatability of ACE estimation.

## IV. Discussion

We hypothesized that averaging FPDCs with weakly correlated oscillation patterns acquired using different harmonic frequencies and steering angles would improve FPDC linearity, thereby improving the robustness of RFM-based ACE in harmonic imaging. This hypothesis was first validated through phantom studies. The results showed that the correlations between FPDCs acquired at different transmission frequencies or steering angles decreased as the frequency or angular separation increased (Figs. 4a-d). Therefore, larger transmit frequency intervals and steering angle increments are generally preferred because they produce greater decorrelation among FPDCs, leading to more effective suppression of oscillations through averaging. However, the available bandwidth of the transducer limits the range of transmit frequencies that can be used. Consequently, only three harmonic transmit frequencies with 0.5 MHz spacing were selected in this study to maximize FPDC decorrelation while remaining within the usable bandwidth of the probe (bandwidth of the probe from 1.8 - 5 MHz). Similarly, for a given $\Delta\theta$, the number of steering angles is constrained by the directivity of the transducer. Therefore, three steering angles were selected as a practical compromise between achieving sufficient FPDC decorrelation and maintaining adequate signal quality.

Additionally, compared to the conventional RFM method, the proposed method improved the linearity of the FPDCs and reduced the CVs of the estimated attenuation coefficients, thereby improving the accuracy and robustness of the mean attenuation coefficient estimation (Figs. 5b, c, e, and f). Additionally, the larger inter-block variations (CVs) observed with the conventional RFM which may be attributable to oscillations in the FPDCs, resulting in inconsistent linear fitting across different analysis blocks. In contrast, the proposed method effectively suppressed these oscillations, resulting in more consistent linear fitting, lower inter-block variation (lower CVs), and higher median $R^2$ values than the conventional RFM across phantoms with different calibrated attenuation coefficients. These improvements were maintained even in the presence of phase aberration introduced by the pork belly layer (Fig. 6), demonstrating the robustness of the proposed approach under more challenging imaging conditions.

Although the ACE obtained using the two methods differed by 0.01-0.06 dB/cm/MHz for *in vivo* study, the lower CV and higher $R^2$ achieved by the proposed method suggest that its ACE are likely to be more accurate and reliable (Fig. 7). Furthermore, the proposed method demonstrated clinical feasibility in a pilot study of 15 patients, achieving a stronger correlation with MRI-PDFF than the conventional RFM (Fig. 8). It also reduced the pooled SD of repeated ACE measurements, indicating improved robustness and repeatability.

Focused beam transmission was used in this study because it provides higher acoustic energy and deeper penetration. However, unfocused wave transmissions (e.g., plane waves or diverging waves) or wide-beam insonification strategies could also be applied with the proposed method, as these approaches have been widely adopted in other ultrasound imaging modalities, including B-mode imaging, shear wave imaging, and ultrasound microvascular imaging [23-30].

On the other hand, increasing the number of data ensembles acquired may further improve the performance of ACE in *in vivo* studies (temporal domain decorrelation). Physiological motion, particularly respiratory motion, introduces temporal decorrelation of the ultrasound speckle pattern, resulting in low-correlated FDPCs. Consequently, acquiring more ensembles over the respiratory cycle increases the likelihood of capturing decorrelated FDPCs, potentially improving the performance of the proposed ACE method. However, respiratory motion is highly subject dependent and cannot be precisely controlled in clinical practice. In general, respiratory motion is expected to be either sufficiently large to completely displace the tissue region being quantified, resulting in highly decorrelated FDPCs, or sufficiently small to introduce only negligible decorrelation. Therefore, although a longer acquisition over the respiratory cycle may improve the performance of RFM by increasing the chance of acquiring decorrelated data, the achievable improvement is expected to subject-dependent and therefore may not be robust in clinical practice. Future studies will investigate adaptive acquisition strategies that determine the optimal number of ensembles based on the measured temporal decorrelation or motion characteristics, thereby maximizing estimation accuracy while maintaining clinically acceptable acquisition times and real-time capability.

Although the proposed method improved attenuation estimation performance in this study, several limitations should be acknowledged. First, the substantial decrease in signal power in highly attenuating media may bias ACE estimation. For example, for an attenuation coefficient of 0.9 dB/cm/MHz and a transmit frequency of 4.5 MHz, the round-trip signal attenuation is approximately 65 dB at a depth of 80 mm. Under such low signal-to-noise ratio (SNR) conditions, the FPDC may deviate from the expected linear decay and become falsely elevated, as demonstrated in [16]. This distorted decay curve can lead to underestimation of the attenuation coefficient estimate (ACE). Therefore, a noise suppression method [16] was incorporated in this study to mitigate this effect; however, its drawback is a reduction in the effective ROI. To further enhance performance, transmission designs with superior penetration [31] or higher power amplifiers for the ultrasound probe [32-34] may be considered in the future.

Secondly, the proposed method was implemented on a CPU-based system and required approximately 47 seconds for beamforming and an additional around 2 seconds for power spectrum estimation and formation of FPDC for each transmission; however, it is well suited for parallel implementation using multi-core computing platforms [35, 36] or GPUs [37-40]. Both the beamforming process and the subsequent post-processing steps are highly parallelizable because the calculations for each transmission of power spectrum and FPDC of different blocks can be performed independently with low data dependency. Therefore, with optimized parallel implementation, the proposed method has strong potential to achieve real-time attenuation coefficient estimation while maintaining the improved estimation accuracy demonstrated in this study.

Another limitation of this study is the relatively small patient cohort used to evaluate the correlation between ACE and MRI-PDFF (n = 15). Although the preliminary results demonstrate a promising correlation with the proposed method, the limited sample size limits the generalizability of the findings. Future studies involving larger and more diverse patient cohorts are needed to further validate the proposed ACE.

Finally, this study assumed a linear frequency dependence on tissue attenuation. However, the frequency dependence of attenuation in biological tissues may exhibit nonlinear behavior [41], Therefore, the proposed method should be extended in future studies to accommodate more general attenuation models and evaluate its performance under nonlinear frequency-dependent attenuation conditions.

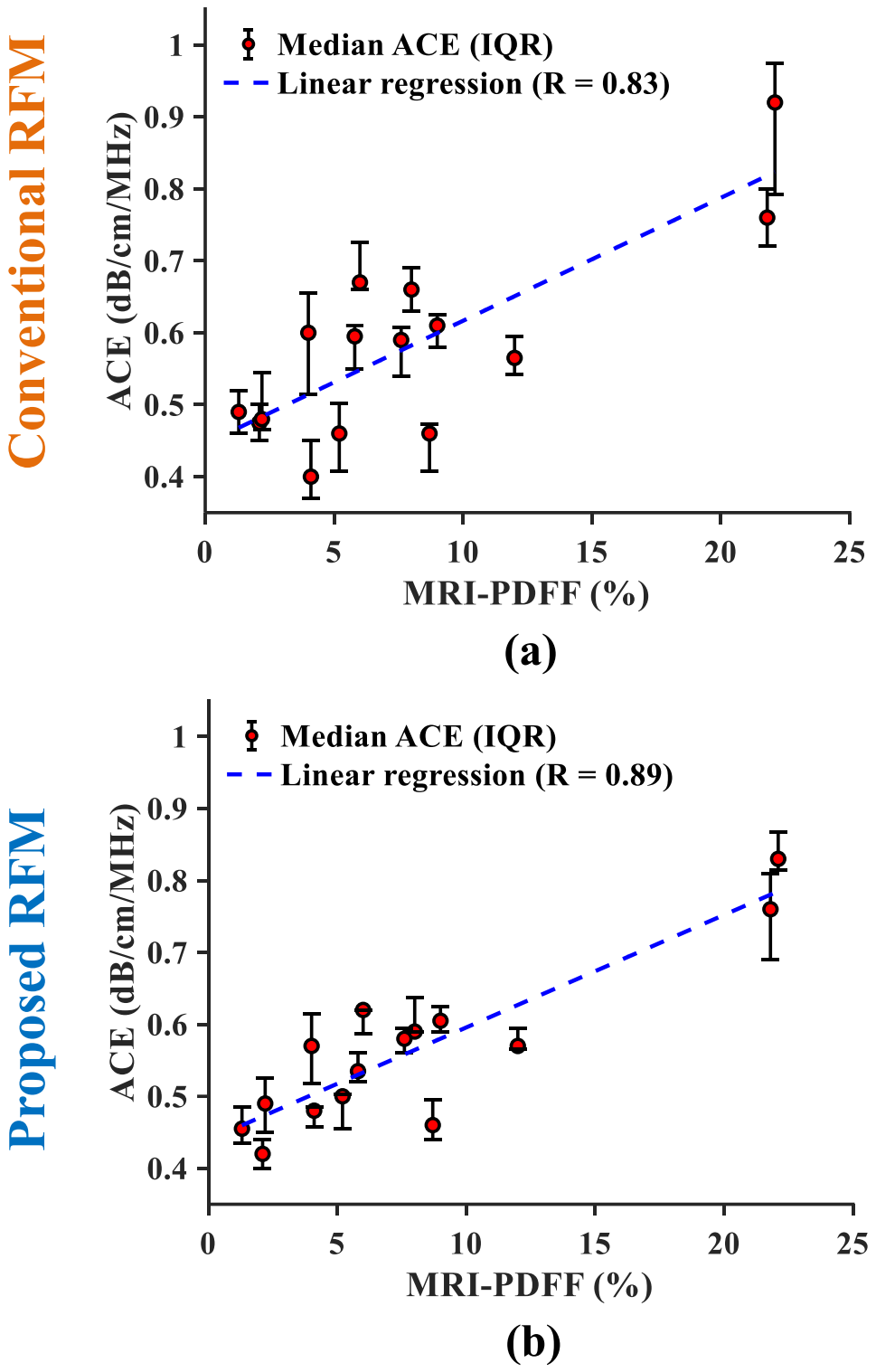

**Fig. 8.** Correlation analysis between median ACE and MRI-PDFF for (a) conventional RFM and (b) proposed RFM methods. The red dot represents the median ACE value from repeated measurements of an individual patient, and error bars indicate the interquartile range (IQR). Blue dashed lines indicate linear regression fits. The proposed RFM method achieved improved correlation with MRI-PDFF (R = 0.89) compared with the conventional RFM method (R = 0.83).

## V. Conclusion

We have proposed the use of multiple transmitted harmonic frequencies and steering angles in RFM. The proposed method provided accurate and more robust attenuation estimation in phantom and *in vivo* liver studies compared to that of the conventional RFM method. The improvement of ACE indicates the feasibility and potential for more robust liver steatosis detection.

## Acknowledgment

This project was supported partially by the National Institutes of Health under award number R01DK127978. The content is solely the responsibility of the authors and does not necessarily represent the official views of the NIH. Mayo Clinic and some authors (U.L, J.Z., and S.C.) have a potential financial interest (Patents/Licensing) related to the technology referenced.